
%
\input harvmac

\def\title#1#2#3{\Title{ESENAT-#1}{#2}\vskip -0.3in
\centerline{{\titlefont#3}}\vskip 0.3in}
\def\title#1#2#3#4{\nopagenumbers\abstractfont\hsize=\hstitle
\rightline{ESENAT-#1}
\rightline{YUMS-#2}%
\bigskip
\vskip 0.7in\centerline{\titlefont #3}\abstractfont\vskip .2in
\centerline{{\titlefont#4}}
\vskip 0.3in\pageno=0}
\def\jsp{\centerline{Jae-Suk Park\footnote{$^{\dagger}$}{e-mail:
jspark@phya.yonsei.ac.kr}}
\bigskip
\centerline{{\it ESENAT Theoretical Physics Group}}
\centerline{{\it Seodamun P.O.~Box \#126, Seoul 120-600, Korea}}
\centerline{{\it and}}
\centerline{{\it Institute for Mathematical Science}}
\centerline{{\it Yonsei University, Seoul 120-749, Korea}}
\bigskip\bigskip}
\def\abs#1#2{\centerline{{\bf Abstract}}\vskip 0.2in
{#2}\Date{#1}}
\def\ack{\bigbreak\bigskip\bigskip\centerline{{\bf Acknowledgements}}\nobreak}

\font\teneufm=eufm10
\font\seveneufm=eufm7
\font\fiveeufm=eufm5
\newfam\eufmfam
\textfont\eufmfam=\teneufm
\scriptfont\eufmfam=\seveneufm
\scriptscriptfont\eufmfam=\fiveeufm
\def\eufm#1{{\fam\eufmfam\relax#1}}

\font\teneusm=eusm10
\font\seveneusm=eusm7
\font\fiveeusm=eusm5
\newfam\eusmfam
\textfont\eusmfam=\teneusm
\scriptfont\eusmfam=\seveneusm
\scriptscriptfont\eusmfam=\fiveeusm
\def\eusm#1{{\fam\eusmfam\relax#1}}

\font\tenmsx=msam10
\font\sevenmsx=msam7
\font\fivemsx=msam5
\font\tenmsy=msbm10
\font\sevenmsy=msbm7
\font\fivemsy=msbm5
\newfam\msafam
\newfam\msbfam
\textfont\msafam=\tenmsx  \scriptfont\msafam=\sevenmsx
  \scriptscriptfont\msafam=\fivemsx
\textfont\msbfam=\tenmsy  \scriptfont\msbfam=\sevenmsy
  \scriptscriptfont\msbfam=\fivemsy

\def\msbm#1{{\fam\msbfam\relax#1}}

\def\a{\alpha}    \def\b{\beta}       \def\c{\chi}       \def\d{\delta}
\def\D{\Delta}    \def\e{\varepsilon}        
               
\def\L{\Lambda}   \def\m{\mu}                 
  \def\o{\omega}      \def\O{\Omega}     \def\p{\psi}
                 
        \def\w{\varphi}    

\def\CA{{\cal A}}

\def\CG{{\cal G}}
\def\CM{{\cal M}}
\def\CN{{\cal N}}
\def\CE{\eusm{E}}
\def\CD{\eusm{D}}
\def\CU{\eusm{U}}
%
\def\rd{\partial}

\def\darr#1{\raise1.5ex\hbox{$\leftrightarrow$}\mkern-16.5mu #1}
\def\Ha{{1\over2}}
\def\ha{{\textstyle{1\over2}}}
\def\fr#1#2{{\textstyle{#1\over#2}}}
\def\Fr#1#2{{#1\over#2}}
\def\tr{\hbox{Tr}\,}

\def\roughly#1{\raise.3ex\hbox{$#1$\kern-.75em\lower1ex\hbox{$\sim$}}}

%
\def\cmp#1#2#3{Comm.\ Math.\ Phys.\ {{\bf #1}} {(#2)} {#3}}
\def\pl#1#2#3{Phys.\ Lett.\ {{\bf #1}} {(#2)} {#3}}
\def\np#1#2#3{Nucl.\ Phys.\ {{\bf #1}} {(#2)} {#3}}
\def\prd#1#2#3{Phys.\ Rev.\ {{\bf #1}} {(#2)} {#3}}

\def\ijmp#1#2#3{Int.\ J.\ Mod.\ Phys.\ {{\bf #1}} {(#2)} {#3}}
\def\mpl#1#2#3{Mod.\ Phys.\ Lett.\ {{\bf #1}} {(#2)} {#3}}
\def\jdg#1#2#3{J.\ Differ.\ Geom.\ {{\bf #1}} {(#2)} {#3}}

\def\top#1#2#3{Topology {{\bf #1}} {(#2)} {#3}}
\def\zp#1#2#3{Z.\ Phys.\ {{\bf #1}} {(#2)} {#3}}
\def\prp#1#2#3{Phys.\ Rep.\ {{\bf #1}} {(#2)} {#3}}
\def\ap#1#2#3{Ann.\ Phys.\ {{\bf #1}} {(#2)} {#3}}
\def\ptrsls#1#2#3{Philos.\ Trans.\  Roy.\ Soc.\ London
{{\bf #1}} {(#2)} {#3}}

\def\am#1#2#3{Ann.\ Math.\ {{\bf #1}} {(#2)} {#3}}

\def\plms#1#2#3{Proc.\ London Math.\ Soc.\ {{\bf #1}} {(#2)} {#3}}
\def\dmj#1#2#3{Duke Math.\  J.\ {{\bf #1}} {(#2)} {#3}}

\def\jgp#1#2#3{J.\ Geom.\ Phys.\ {{\bf #1}} {(#2)} {#3}}

\def\pr{\prime}
\def\ppr{{\prime\prime}}

\def\tw{\tilde{\o}}
\def\bs{{\bf s}}
\def\bbs{{\bf\bar s}}
\def\Da{d_{\!A}}
\def\Dp{\rd_{\!A}}

\def\Dpp{\bar\rd_{\!A}}

\def\gp{\eufm{g}_{\raise-.1ex\hbox{${}_E$}}}
\def\gpc{\eufm{g}_{\raise-.1ex\hbox{${}_E$}}^C}
\def\BZ{\msbm{Z}}
\def\CP{\msbm{CP}}
\def\dw{\d_{{}_W}}
\def\lin#1{\bigskip\leftline{$\underline{\hbox{#1}}$}\bigskip}
\lref\WittenA{
E.\ Witten,
\cmp{117}{1988}{353}.
}
\lref\WittenB{
E.\ Witten,
\ijmp{A 6}{1991}{2273}.
}
\lref\WittenC{
E.\ Witten,
\cmp{141}{1991}{153}.
}
\lref\WittenD{
E.\ Witten,
\jgp{9}{1992}{303}.
}
\lref\WittenE{
E.\ Witten,
\np{B 371}{1992}{191}.
}
\lref\Donaldson{
S.K.\ Donaldson,
\jdg{26}{1987}{141};
\top{29}{1990}{257}.
}
\lref\TKD{
M.\ Thaddeus,
\jdg{35}{1992}{131}\semi
F.\ Kirwan,
J.\ Amer.\ Math.\ Soc. {\bf 5} (1992) 853\semi
S.K.\ Donaldson,
Gluing techniques in the cohomology of moduli spaces,
to appear in Proceedings of the Andreas Foler memorial
volume, eds.\ H.\ Hofer and E.\ Zehnder.
}
\lref\DK{
S.K.\ Donaldson and P.B.\ Kronheimer,
The geometry of four-manifolds (Oxford University Press, New York, 1990)
}
\lref\NSDUY{
M.S.\ Narasimhan and C.S.\ Seshadri,
\am{65}{1965}{540}\semi
S.K.\ Donaldson,
\plms{30}{1985}{1}\semi
K.K.\ Uhlenbeck and S.-T.\ Yau,
Commun.\ Pure \&  Appl.\  Math.\
{\bf 39} (1986) S257; Correction:
Commun.\ Pure \&  Appl.\  Math.\
{\bf 42} (1987) 703\semi
S.K.\ Donaldson,
\dmj{54}{1987}{231}.
}
\lref\AB{
M.F.\ Atiyah and R.\ Bott,
\ptrsls{A 308}{1982}{523}.
}
\lref\Park{
J.-S.\ Park, $N=2$ topological Yang-Mills theory on compact
K\"{a}hler surfaces, Comm.\ Math.\ Phys.\ (1994) to appear.
}
\lref\REPORT{
D.\ Birminham, M.\ Blau, M.\ Rakowski and G.\ Thomson,
\prp{209}{1991}{129}.
}
\lref\AHS{
M.F.\ Atiyah, N.J.\ Hitchin and I.M.\ Singer,
\ptrsls{A 362}{1978}{425}.
}
\lref\Itoh{
M.\ Itoh,
Proc.\ Japan Acad.\ {\bf 59} (1983) 431;
Publ.\ RIMS.\ Kyoto Univ.\ {\bf 19} (1983) 15;
Osaka J.\ Math.\ {\bf 22} (1985) 845;
J.\ Math.\ Soc.\ Japan {\bf 40} (1988) 9.
}
\lref\Kim{
H.J.\ Kim,
Math.\ Z.\ {\bf 195} (1987) 143.
}
\lref\Kobayashi{
S.\ Kobayashi,
Differential geometry of complex vector bundle
(Princeton University Press, Princeton, 1987)
}
\lref\DH{
J.J.\ Duistermaat and G.J.\ Heckmann,
Invent Math.\ {\bf 69} (1982) 259.
}
\lref\Migdal{
A.\ Migdal, Zh.\ Eksp.\ Teor.\ Fiz.\ {\bf 69} (1975) 810 (Sov.\ Phys.\
Jetp.\ {\bf 42}, 413).
}
\lref\Thompson{
G.\ Thompson,
Topological gauge theory and Yang-Mills theory,
Lecture notes of Trieste summer school, June 1992.
}
\lref\HP{
S.~Hyun and J.-S.~Park, in progress.
}

\lref\tdymt{
H.G.\ Dosch and V.F.\ Muller,
Fortscher. Phys. {\bf 27} (1979) 547\semi
V.\ Kazakov and I.\ Kostov,
\pl{B 105}{1981}{453}\semi
L.\ Gross, C.\ King and A.\ Sengupta,
\ap{194}{1989}\semi
B.K.\ Driver,
\cmp{123}{1989}{575}\semi
B.\ Rusakov,
\mpl{A 5}{1990}{693}\semi
D.\ Fine,
\cmp{134}{1990}{273}; {\bf 140} (1991) 321\semi
M.\ Blau and G.\ Thompson,
\ijmp{A 7}{1992}{3781}\semi
R.\ Formann,
\cmp{151}{1993}{39}.
}
\lref\tymt{
L.\ Baulieu and I.M.\ Singer,
\np{(Proc.\ Suppl.) 5B}{1988}{12}\semi
J.M.F.\ Labastida and M.\ Pernici,
\pl{B 212}{1988}{56}\semi
R.\ Brooks, D.\ Montano and J.\ Sonnenschein,
\pl{B 214}{1988}{91}\semi
D.\ Birmingham, M.\ Rakowski and G.\ Thompson,
\np{B 315}{1989}{577}\semi
S.\ Ouvry and G.\ Thompson,
\np{B 344}{1990}{371}\semi
A.\ Galperin and O.\ Ogievetsky,
\cmp{139}{1991}{377}.
}
\lref\Kanno{
H.\ Kanno,
\zp{C 43}{1989}{477}\semi
S.\ Ouvry, R.\ Stora and P.\ van Ball,
\pl{B 220}{1989}{1590}.
}
\lref\AJ{
M.F.\ Atiyah and L.C.\ Jeffrey,
\jgp{7}{1990}{120}\semi
M.\ Blau and G.\ Thompson,
\cmp{152}{1993}{41}\semi
J.\ Kalkman,
\cmp{163}{1993}{447}.
}
\lref\Kirwan{F.\ Kirwan, Cohomology of quotients in symplectic
and algebraic geometry (Princeton Univ.~Press, 1987).
}
\lref\Gross{D.J.\ Gross, \np{B 400}{1993}{161}\semi
D.J.\ Gross and W.\ Taylor, \np{B 400}{1993}{181};
{\bf B 403} (1993) 395 \semi
J.A.\ Minahan, \prd{D 47}{1993}{3430}.
}
\lref\FU{
D.S.\ Freed and K.K.\ Uhlenbeck, Instantons and four-manifolds
(Springer-Verlag, New York, 1991)
}
\lref\NN{
A.\ Newlander and L.\ Nirenberg, \am{65}{1957}{391}.
}
\lref\MW{
J.\ Marsden and A.D.\ Weistein, Rep.\ Math.\ Phys.\ {\bf 5} (1974) 121.
}
\lref\GH{
P.~Griffiths and J.~Harris,
Principles of algebraic geometry
(John Wiley \&\ Sons Inc., 1978)
}

\title{93-02}{93-11}{Holomorphic Yang-Mills Theory}
{on  Compact  K\"{a}hler Manifolds}
\bigskip
\bigskip
\jsp
\bigskip
\abs{May, 1993; February, 1994}{
We propose $N=2$ holomorphic Yang-Mills theory on compact K\"{a}hler
manifolds
and show that there exists a simple mapping
from the $N=2$ topological Yang-Mills theory.
It follows that intersection parings on the moduli
space of Einstein-Hermitian connections can be determined
by examining the small coupling behavior of the $N=2$ holomorphic Yang-Mills
theory. This paper is a higher dimensional generalization of the Witten's
work on physical Yang-Mills theory in two dimensions.
\bigskip
}


\newsec{Introduction}

Few years ago, Witten introduced the topological Yang-Mills (TYM)
theory \WittenA\ to give a quantum field theoretical interpretation of
the Donaldson polynomial invariants of smooth four-manifolds \Donaldson\DK.
The construction of the Lagrangian and the various mathematical
structures of the theory are  understood well enough
\tymt\Kanno\AJ\REPORT. However, the TYM theory has not provided
any new insight to the explicit computations of the invariants.

Recently, we proposed a $N=2$ supersymmetric TYM ($N=2$ TYM)
theory on compact K\"{a}hler surfaces \Park. This theory can be easily
generalized to arbitrary dimensional compact K\"{a}hler manifolds,
which is a field theoretical interpretation
of intersection parings on
the moduli space of stable bundles.
In this paper, we propose $N=2$ holomorphic Yang-Mills (HYM) theory on
compact K\"{a}hler manifolds. Our main result is a proof of the
equivalence of this theory to the $N=2$ TYM theory.
The $N=2$ HYM theory may enhance the computability of the invariants.

This paper
is a generalization of the Witten's work
in two dimensions \WittenD. He showed that the TYM theory is equivalent
to physical Yang-Mills (YM) theory in two dimensions. Since
physical YM theory can be exactly solved in two dimensions
\Migdal\tdymt\WittenC, he was able to obtain  general
expressions for the intersection parings on the moduli space of
flat connections\foot{Similar result for $SO(3)$ case was obtained
independently in \TKD.}.

Classically, the $N=2$ HYM theory is equivalent to physical YM theory
after restricting the space $\CA$ of all connections to its subspace
$\CA^{1,1}$ having curvature only of type $(1,1)$.
In sect.~2, we discuss few aspects on the physical YM
theory restricted to $\CA^{1,1}$ (which will be called HYM theory).
We compare the HYM theory with physical YM theory in two
dimensions.
In sect.~3, we formulate the $N=2$ HYM
theory. After a brief review on the $N=2$ TYM theory in sect.~3.1,
we show that there exists a simple mapping to the $N=2$ HYM theory.
This ensures that the $N=2$ HYM theory is well-defined as a quantum
field theory.  We also show that the partition function of the $N=2$
HYM theory can
be expressed as the sum of the contributions of the critical points
which can be determined by some differential-topological methods.
It follows that the intersection parings on the moduli space of
stable bundles can be determined by solving the $N=2$ HYM theory.
Finally, we give some remarks in sect.~4.

The basic idea of this paper was announced
previously in the last section of \Park.
While this work is in progress, we received a lecture
notes by Thompson \Thompson\ who proposed a similar model but in
a different language for compact K\"{a}hler surfaces.

\newsec{Holomorphic Yang-Mills Theory}

Let $(M,\o)$ be a complex $n$-dimensional compact
K\"{a}hler manifold\foot{
Throughout this section we refer freely to \DK\Kobayashi\
for mathematical details.} with K\"{a}hler form $\o$.
Let $E$ be a complex vector bundle over $M$ with a reduction of
structure group to $SU(r)$.
Let $\eufm{su}(r)$ be the Lie algebra of $SU(r)$.
We write $\gp = E\times_{Ad}\eufm{su}(r)$ for the Lie algebra
bundle associated with $E$ by adjoint representation.
Let $\CA$ be the space of all
connections on $E$ and $\CG$ be the group of gauge
transformations.  We introduce a positive definite quadratic form
$(a,b) = -\tr ab$ on $\eufm{su}(r)$, where $\tr$
denotes the trace in the $r$ dimensional representation.
The action functional of physical YM theory
is given by
\eqn\james{
S(A) = -\Fr{1}{8\pi^2\e}\int_M \tr F_{\!A}\wedge * F_{\!A},
}
where $\e$ is a non-negative constant. An YM connection is
a critical point $\Da ^* F_{\!A} =0$ of the YM action.
If we use the Bianchi identity $\Da  F_{\!A} =0$ and the K\"{a}hler
identity,
\eqn\kahler{
\Dpp^* = i[\Dp,\L], \qquad \Dp^* = -i[\Dpp,\L],
}
the YM equation can be written as \ItohB
\eqn\afd{\eqalign{
-in\Dpp f +  \Dpp^* F_{\!A}^{0,2}&=0,\cr
in\Dp f +  \Dp^* F_{\!A}^{2,0}&=0,\cr
}
}
where  $f = \fr{1}{n}\L F_{\! A}$.

For $n=1$, the YM action becomes
\eqn\ooa{
S(A) = -\Fr{1}{8\pi^2\e}\int_M \o\tr(f^2).
}
This does not depend on metric but only on the cohomology
class of the K\"{a}hler form (the volume form). Clearly, the YM
action is minimized by a flat connection. The equation of motion
$\Da f =0$ shows that any YM connection is either flat $(f=0)$
or reducible $(f\neq 0)$.
For $n=2$, it is well-known
that the YM action can be decomposed as
\eqn\kaon{
S(A,k) = -\Fr{1}{4\pi^2\e}\int_M \tr F^+_A\wedge * F^+_A
+\Fr{k}{\e},
}
where
\eqn\kke{
k = \int_M c_2(E) = \Fr{1}{8\pi^2}\int_M \tr F_{\!A}\wedge F_{\!A}
\in \msbm{Z}.
}
The positivity of the instanton number $k$ is a topological
restriction of $E$ in order to admit an ASD connection. For $k > 0$,
the YM action is minimized by the ASD connections.

We propose a simple higher dimensional analogue of physical
YM theory on Riemann surfaces. The basic idea is to {\it restrict}
$\CA$ to $\CA^{1,1}$. Then, the YM action becomes
\eqn\joan{\eqalign{
I(A,c_2)
&=-\Fr{1}{8\pi^2\e}\int_M \tr F^{1,1}_A\wedge * F^{1,1}_A
\cr
&= -\Fr{n^2}{8\pi^2\e}\int_M \Fr{\o^n}{n!}\tr f^2 + \Fr{1}{8\pi^2\e}
\int_M \tr (F^{1,1}_A\wedge F^{1,1}_A)\wedge \Fr{\o^{n-2}}{(n-2)!}.
\cr
}}
We may call this theory holomorphic Yang-Mills (HYM) theory,
since each connection $A$ lies in $\CA^{1,1}$, i.e.\
$F^{0,2}_{\!A} = \Dpp^2 =0$, and each operator $\Dpp$ defines a
holomorphic structure $\CE_A$ on $E$ by the integrability theorem of
Newlander-Nirenberg \NN. The action functional of
the HYM theory is minimized by
an integrable connection $A$ satisfying $f = 0$
if and only if
the following inequality of L\"{u}bke holds;
\eqn\lubke{
c_2 := \Fr{1}{8\pi^2}
\int_M \tr(F^{1,1}_A\wedge F^{1,1}_A)\wedge \Fr{\o^{n-2}}{(n-2)!}\geq 0.
}
The equality holds if and only if the connection is flat. We will assume
$c_2 > 0$ throughout this paper.
Let $\CE$ be a holomorphic vector bundle over $M$,
then an unitary connection $A$ of $\CE$ is Einstein-Hermitian
if $\L F_A = nf$ is a constant
multiple of the identity endomorphism. Using the Chern-Weil theory,
the constant can be shown to be proportional to
deg$(\CE) = \int_M c_1(\CE)\wedge \Fr{\o^{n-1}}{(n-1)!}$.
Note that an $SU(r)$ connection $A \in \CA^{1,1}$ satisfying $f =0$
endows $E$ with a holomorphic vector bundle $\CE_A$ satisfying
$$
c_1(\CE_A) =0,\qquad  \int_M c_2(\CE_A)\wedge
\Fr{\o^{n-2}}{(n-2)!} = c_2,
$$
and defines an Einstein-Hermitian structure on $\CE_A$.
In this sense,
a connection  $A\in\CA^{1,1}$ on $E$ is Einstein-Hermitian (EH)
or Hermitian Yang-Mills if $f =0$.
That is, the action of the HYM theory is minimized by the EH connections
provided that the inequality \lubke\ holds. We denotes $\CM$
the moduli space of EH connections.
One can also define the moduli space of holomorphic vector bundles
as the isomorphism class of $\CE_A$'s. It turns out that this moduli
space can be identified with the complex quotient,
$\CA^{1,1}/\CG^\msbm{C}$,
where $\CG^\msbm{C}$ denotes the complexification of $\CG$.
Due to the theorem of Donaldson-Uhlenbeck-Yau \NSDUY,
the moduli space $\CM^*$ of irreducible
EH connections is diffeomorphic to the moduli space
$\CA^{1,1}_S/\CG^\msbm{C}$ of
$\o$-stable bundles .

\lin{Comparison With the Two Dimensional Case}

The YM theory restricted to $\CA^{1,1}$
has great similarity with physical YM theory
on Riemann surfaces \WittenC\WittenD.
First of all, the action functional depends only on
the cohomology class of $\o$ and $c_2(E)$ in  $H^{1,1}(M)$.
And, any Yang-Mills connection
$A\in \CA^{1,1}$ (HYM connection) is either EH $(f=0)$  or
reducible $(f\neq 0)$
\eqn\ffd{
i(\Dpp -\Dp)f =0 \rightarrow \Da  f =0.
}
The HYM theory has global scaling invariance. If we scale the
K\"{a}hler form by $\o \rightarrow \o/t$ for any positive
real number $t$, the action is invariant under $\e \rightarrow
t^{2-n} \e$.  Note that the coupling constant is scaling invariant
only for $n=2$. Finally, the action functional HYM theory is a norm
squared of moment map up to topological term $c_2$.
This is the crucial property
of physical YM theory on Riemann surfaces, which led Witten
to determine the general expressions of the intersection pairings
on the moduli space of flat connections.

Note that $\CA$
inherits the complex structure and the K\"{a}hler structure of
$M$.  For the given complex structure $J$
on $M$, we can introduce a complex structure on $\CA$ by identifying
$T^{1,0}\CA$ and $T^{0,1}\CA$, in the decomposition of tangent
space $T \CA = T^{1,0}\CA \oplus T^{0,1}\CA$, with
$\gp$-valued
(1,0)-forms and (0,1)-forms on $M$ respectively. Then, $\CA$ has
natural K\"{a}hler structure,
\eqn\aac{
\tw = \Fr{1}{4\pi^2}\int_M \tr(\d A^\pr\wedge \d A^\ppr)
\wedge\Fr{\o^{n-1}}{(n-1)!},
}
where $\d A^\pr\in \O^{1,0}(\gp)$ and $\d A^\ppr\in \O^{0,1}(\gp)$.
And, $\CG$ acts on $\CA$ by isometries.
If we {\it restrict} $\CA$ to its subspace $\CA^{1,1}$,
the subspace $\CA^{1,1}$ is
preserved by the action of $\CG$ and its smooth
part has the K\"{a}hler structure given
above with $\Dp\d A^\pr=\Dpp\d A^\ppr=0$.
Let $Lie(\CG)$ be the Lie algebra of $\CG$, which can be
identified with the space of $\gp$-valued zero-form.
Then, we have a moment map,
$\eufm{m}:\CA^{1,1}\rightarrow \O^0({\gp})^*$,
\eqn\aad{
\eufm{m}(A) = -\Fr{1}{4\pi^2}F^{1,1}_{\!A}\wedge\Fr{\o^{n-1}}{(n-1)!}
= -f\Fr{n}{4\pi^2}\Fr{\o^{n}}{n!},
}
where $\O^0(\gp)^*=\O^{2n}(\gp)$ denotes dual of $\O^0(\gp)$.
Thus, the action functional of the HYM theory is the norm squared of
the moment map, $(\eufm{m},\eufm{m}) \approx -\int_M \fr{\o^n}{n!}
\tr f^2$, with respect to a metric on $Lie(\CG)$ determined
by the measure $\o^n/n!$ of $M$.
The reduced phase space
(symplectic quotient) $\eufm{m}^{-1}(0)/\CG$
can be identified with the moduli space $\CM$ of EH connections.
The reduced phase space also has the K\"{a}hler structure descended
{\it symplectically}\foot{The K\"{a}hler structure does not
descend to $\CA/\CG$ in general.} from $\CA^{1,1}$ by
the reduction theorem of Mardsden-Weinstein \MW.

\lin{Predictions of the Non-Abelian Localization Theorem}

We can {\it formally} define the partition function of the
HYM theory by
\eqn\vfa{\eqalign{
Z(\e,c_2)
&= \exp(-\Fr{c_2}{\e})\times Z(\e)\cr
&= \exp(-\Fr{c_2}{\e})\times\Fr{1}{\hbox{vol}(\CG)}\int_{\CA^{1,1}}
\CD\! A \exp\biggl[
\Fr{n^2}{8\pi^2\e}\int_M\!\Fr{\o^n}{n!} \tr f^2\biggr].\cr
}}
According to the Non-Abelian localization theorem of Witten \WittenD, this
partition function can be expressed as sum of the contributions
of the critical points:
\eqn\ooc{
Z(\e,c_2) = \exp(-\Fr{c_2}{\e})\times \sum_{\a\in \CN} Z_\a(\e),
}
where $\CN$ denotes the moduli space of the HYM connections
of given topological type.

Due to the great similarity with the physical YM theory
on Riemann surfaces, we can repeats many of the manipulations given
by Witten.  For example, we can introduce
a first order formalism of the HYM theory defined by the action \WittenC,
\eqn\ahym{\eqalign{
I(A,\w, c_2) = &-\Fr{1}{4\pi^2}\int_M \tr\bigl(i\w F^{1,1})\wedge
\Fr{\o^{n-1}}{(n-1)!}
-\Fr{\e}{8\pi^2}\int_M \Fr{\o^n}{n!}\,\tr\w^2\cr
&+\Fr{1}{8\pi^2\e}\int_M \tr(F^{1,1}\wedge
F^{1,1})\wedge \Fr{\o^{n-2}}{(n-2)!},\cr
}}
which is equivalent to the original one after integrating
$\w \in \O^0(\gp)$ out. This formalism is
useful for studying the zero coupling limit of
$Z(\e)$, such that the only contribution to
the path integral comes from the reduced phase space:
\eqn\zkz{
Z(0)
=\Fr{1}{\hbox{vol}(\CG)}\int_{\CA^{1,1}}
\!\!\CD\! A\,\CD\w \exp\biggl(
\Fr{in}{4\pi^2}\int_M \Fr{\o^n}{n!}\tr\w f
\biggr).
}
Let $\# Z$ be the number of the center $Z(SU(r))$ of $SU(r)$
which coincides to $Z(\CG)$.
If there are no reducible EH connections such that
$\widetilde\CG=\CG/Z(SU(r))$ acts freely
on the moduli space $\CM$, one can  expect that $Z(0)$ is
related to the volume of the moduli space.
One can rewrite the partition function
\zkz\ as
\eqn\kkz{
Z(0)
=\Fr{1}{\# Z}\int_{\CA^{1,1}/\widetilde\CG}
\!\!\!\!\CD\! A^\pr\,\CD\w\,\exp\biggl(
\Fr{in}{4\pi^2}\int_M \Fr{\o^n}{n!} \tr\w f
\biggr),
}
where
the measure $\CD\! A^\pr$ denotes quotient measure  on
$\CA^{1,1}/\widetilde\CG$. The quotient measure
can be constructed by adopting  the Faddev-Popov-BRST procedure.
One can calculate $Z(0)$ by following the procedures in sect.\ $2.2$ of
\WittenC.  The linearization of local equations which cut out
the moduli space $\CM$ of EH connections insides $\CA^{1,1}$
can be written as
\eqn\qqw{
Q B = 0,\quad\hbox{where}\quad Q = P^{1,1}_+ \Da \oplus \Da^*:\O^{1}(\gp)
\rightarrow \O^{1,1}_+(\gp)\oplus \O^0(\gp),
}
where $P^{1,1}_+$ is the projection operator, acting on
$\O^2(\gp)$, to the space $\O^{1,1}_+(\gp)$ of $(1,1)$-form
parallel to the K\"{a}her form.
If we assume that there is an isolated irreducible
EH connection $A$, we will end up with the following determinant ratio:
\eqn\krd{
\Fr{1}{\#Z}\Fr{\hbox{det}\,\D_0}
{|\hbox{det}\, Q|},
}
where  $\D_0 = \Da^* \Da$ denotes the Laplacian acting
on zero-forms. Clearly, the determinant $\hbox{det}\,\D_0$
is given by the Gaussian integral over the standard ghost
and anti-ghost. One can find that
\eqn\krr{
|\hbox{det}\, Q| =\sqrt{Q Q^*}
 =\sqrt{\hbox{det}\,\D_0 \hbox{det}\,\D_{1,1}^+},
}
where $Q^*$ is the adjoint of $Q$ and
$\D_{1,1}^+ =(P^{1,1}_+\Da)(P^{1,1}_+\Da)^*$ is the Laplacian
acting on $\O^{1,1}_+(\gp)$. Thus, eq.\krd\ becomes
\eqn\krf{
\Fr{1}{\#Z}\sqrt{
\Fr{\hbox{det}\,\D_0}
{\hbox{det}\, \D_{1,1}^+}},
}
This is trivial due to the isomorphism
(specific to compact K\"{a}hler manifold)
between $\O^{0}(\gp)$ and $\O^{1,1}_+(\gp)$
by the formula
\eqn\formu{
\O^{1,1}_+(\gp) = \O^0(\gp)\otimes \o.
}
If the moduli space is a smooth manifold, one can show that
$Z(0)$ reduces to the volume of $\CM$ with the trivial correction factor
$1/\# Z$,
\eqn\tyh{
Z(0) =\Fr{1}{\# Z}\hbox{vol}(\CM).
}
This should be viewed as only a formal result since
the moduli space is rarely compact.
Note that we assumed that there are no reducible EH connections.
This is necessary to ensure a well-defined determinant
ratio \krd. It is, however, not sufficient condition (except $n=1$
case) because there is another obstruction to  have a smooth
moduli space. This important issue will be discussed in the next section.

\newsec{Mapping from $N=2$ TYM theory}

In the previous section, we defined the HYM theory as
the restriction of physical Yang-Mills theory to $\CA^{1,1}$.
However, the treatments in that section were essentially classical.
The question is whether the theory is well defined as a quantum field
theory. In particular, we should impose certain constraints
which restricts $\CA$ to $\CA^{1,1}$ in such a way that no non-trivial
quantum correction appears.
Another important issue is to ensure the finiteness
of the theory. In this section, we construct
a $N=2$ supersymmetric HYM ($N=2$ HYM) theory which can be shown to be
well-defined as a quantum field theory.
Our arguments will be based on a simple
mapping from the $N=2$ TYM theory \Park.
This is analogous
to the mapping, discovered by Witten, from the $N=1$ TYM theory  to physical
YM theory in two dimensions \WittenD.
The equivalence of the $N=2$ HYM theory
to the $N=2$ TYM theory will enable us to interpret the $N=2$ HYM theory
as a simple field theoretical model of the Donaldson invariants of compact
K\"{a}hler surface and their cousins in higher dimensions.

\subsec{N=2 TYM theory}

The purpose of this subsection is to sketch the $N=2$ TYM theory very
rapidly \Park. At this time, we formulate $N=2$ TYM theory on an arbitrary
dimensional compact K\"{a}hler manifold.
The basic multiplet of the theory is
the $[A,(\p,\bar\p),\w]$ with the $N=2$ transformation law
\eqn\symphony{\eqalign{
&\bs A^\pr =-\p,\cr
&\bbs A^\pr =0,\cr
&\bs A^\ppr =0,\cr
&\bbs A^\ppr =-\bar\p,\cr
}\qquad\eqalign{
&\bs \p = 0,\cr
&\bbs\p = -i\Dp\w,\cr
&\bs\bar\p = -i\Dpp\w,\cr
&\bbs \bar\p = 0,\cr
}\qquad\eqalign{
\bbs\w=0,\cr
\bs\w=0,\cr
}}
where $\p \in \O^{1,0}(\gp)$, $\bar\p \in \O^{0,1}$ and $\w \in
\O^0(\gp)$.
This algebra can be obtained by decomposing the original
$N=1$ algebra of ref.\ \WittenA\ according to the complex structures
on $M$ and $\CA$.
The commutation relations of the fermionic symmetry generators
$\bs,\bbs$ are
\eqn\kkg{
\bs^2=0,\qquad (\bs\bbs +\bbs\bs) = i\Da \w = -i\d_\w,\qquad \bbs^2=0,
}
where $\d_\w$ is the generator of a gauge transformation with
infinitesimal parameter $\w$.  We introduce two ghost numbers
$(U,R)$, which assign $(1,1)$ to $\bs$ and $(1,-1)$ to $\bbs$.
To write a topological action, it is sufficient to
introduce an anti-ghost $B\in \O^2(\gp)$
in the adjoint representation with
$(U,R) = (-2,0)$.
Then \kkg\ naturally leads us to
multiplet $[B,(i\c, -i\bar\c), H]$ with
transformation law
\eqn\mazuruka{\eqalign{
&\bs B = -i\c,\cr
&\bbs B =i\bar\c,\cr
&\bs\bar\c = H -\fr{1}{2}[\w,B],\cr
&\bbs\c =H+\fr{1}{2}[\w,B],\cr
}\qquad
\eqalign{
&\bs\c = 0,\cr
&\bbs\bar\c = 0,\cr
&\bs H =-\fr{i}{2}[\w,\c],\cr
&\bbs H =-\fr{i}{2}[\w,\bar\c].\cr
}}
We assume that $B,\c,\bar c,H \in \O^2_+(\gp)$, i.e.
$B= B^{2,0} + B^{0,2} + B^0\o$.
We further assume that  $\c^{2,0}$ and $\bar\c^{0,2}$ to
satisfy $\Dp\c^{2,0} =\Dpp\bar\c^{0,2} =0$ if $n >2$.

The action of $N=2$ TYM theory can be written in the form
\eqn\zzi{\eqalign{
S
= \Fr{\bs\bbs-\bs\bbs}{2} V
=\Fr{\bs\bbs-\bbs\bs}{2}\left(-\Fr{1}{h^2}\int_M \tr B\wedge * F\right).
}}
Note that  $V$ has $(U,R)=(-2,0)$, such that the action
has $(U,R)=(0,0)$
\foot{One can add $-\fr{1}{h^2}\int_M \tr \c\wedge *\bar\c$
to $V$. The resulting theory
will be most complete but equivalent to that considered here \Park.
Our choice leads to much
simpler arguments in the next subsection. I would like to thank G.\
Thompson for pointing this out.}.
We find that
\eqn\gorby{\eqalign{
S =
&\Fr{1}{h^2}\!\int_M\!\!\! \tr\biggl[
  -iH^{2,0}\wedge*F^{0,2}
  - iH^{0,2}\wedge*F^{2,0}
  +i\c^{2,0}\wedge *\Dpp\bar\p
  +i\bar\c^{0,2}\wedge *\Dp\p
  \phantom{\biggl]}
  \cr
& -\fr{i}{2}[\w,B^{2,0}]\wedge * F^{0,2}
  +\fr{i}{2}[\w,B^{0,2}]\wedge * F^{2,0}
  -\biggl( in H^{0}f -i\bar\c^{0}\L \Dpp\p
  \phantom{\biggl]}
  \cr
& -i\c^{0}\L \Dp\bar\p
  +\ha B^{0}\L\left((i\Dp \Dpp - i\Dpp\Dp)\w -2[\p,\bar\p]\right)
  \biggr)\Fr{\o^n}{n!}\biggr]\;,
  \cr
}}
where we have used $\Dpp\Dpp\w = [F^{0,2},\w]$ and $\Dp\Dp\w =[F^{2,0},\w]$.
The integration over  $H$ gives a delta function
$\d(F^+ =0)$. The transformation law \symphony\
shows that $\w$ is a covariant constant, $\Da\w =0$, at the
fixed point locus. Thus,
the path integral reduces to an integral over
the moduli space of EH
connections and the space of $\w$ zero-modes.
If there are no reducible EH connection, the path integral
is localized to an integral over the moduli space $\CM^*$
of irreducible EH connections.

{}From \gorby,
one can see the zero-modes  of the fermionic variables
$\p$ and $\bar\p$ are given by
\eqn\qqa{
\Dp\p =\Dp^*\p =0,\qquad
\Dpp\bar\p=\Dpp^*\bar\p=0.
}
The $\p$ and $\bar\p$ zero-modes represent the holomorphic
and the anti-holomorphic (co-)tangent vectors on $\CM$ at $A$
respectively, provided that its neighborhood $[A]$ in $\CM$ is smooth.
The zero-modes of $\c$ and $\bar\c$ satisfy
\eqn\qqb{
\biggl\{
\eqalign{
i\Dp\c^{0} +\Dp^* \c^{2,0}  &=0,\cr
-i\Dpp\bar\c^{0} + \Dpp^* \bar\c^{0,2} &=0,\cr
}\Longrightarrow\biggl\{
\eqalign{
\Dp\c^{0}=\Dp^* \c^{2,0}  &=0,\cr
\Dpp\bar\c^{0} = \Dpp^* \bar\c^{0,2} &=0.\cr
}}
Thus, there are no $\c^0,\bar\c^0$ zero-modes if every EH connection  is
irreducible. Note that the zero-mode of $\bar\c^{2,0}$
satisfies $\Dpp^* \bar\c^{0,2}=  \Dpp \bar\c^{0,2}=0$.
Let $\CE_A$ be the holomorphic structure induced by an
EH connection $A$ and $\hbox{End}_0(\CE_A)$ be the trace-free
endomorphisms of $\CE_A$. Then a zero-mode $\bar\c^{0,2}$
is an element of $H^2(\hbox{End}_0(\CE_A))$.
The moduli space $\CM^*$ of irreducible EH connections is
smooth at $[A]$ if $H^2(\hbox{End}_0(\CE_A))
=0$ \AHS\Itoh\Kim\Kobayashi.
Thus, the moduli space $\CM$ is a smooth K\"{a}hler manifold
if there are no $\c$ and $\bar\c$ zero-modes, and
its complex dimension is identical to the number of
$\bar\p$ zero-modes. An important point to note is that the theory
has the $U$ number anomaly equals to $2(\#\bar\p -\#\bar\c^{2,0} -
\#\bar\c^0)$ where $\#$ denotes the number of zero-modes.
On the other hand, the $R$ number anomaly is absent.

The set of topological observables can be easily constructed.
And, by standard analysis of cohomological
theory, one can show that their expectation values are
topological invariants in general circumstance \WittenA\Park.
This is valid only at formal level because the contributions
of singularities and non-compactness can make the topological
interpretations of those path integrals difficult \WittenB.
To the author's knowledge, only the existence of the EH connections
are established for $n > 2$ cases. However, we will
proceed our analysis assuming favorable situations.
It is sufficient, for our purpose,
to consider the following observables;
\eqn\aai{\eqalign{
&W^{2,2} =\Theta= \Fr{1}{8\pi^2}\int_M \tr(\w^2)\wedge \Fr{\o^n}{n!},\cr
&W^{1,1} =\tw =
\Fr{1}{4\pi^2}\int_M \tr(i\w F^{1,1} +\p\wedge\bar\p)\wedge
\Fr{\o^{n-1}}{(n-1)!},\cr
&W^{0,0} =-\tilde c_2= -\Fr{1}{4\pi^2}\int_M \tr(F^{2,0}\wedge F^{0,2}
+\Ha F^{1,1}\wedge F^{1,1})\wedge\Fr{\o^{n-2}}{(n-2)!},\cr
}}
which are both $\bs$ and $\bbs$ invariant.
A $W^{m,m}$ carrying $(U,R) = (2m,0)$
is an element of $(m,m)$-th Dolbeault cohomology group on $\CA/\CG$,
which depends only on the cohomology class of $\o$ (and $c_2(E)$
for $m=0$).

Now we assume that there are no $\c$ and $\bar\c$ zero-modes,
such that the path integral is localized to the moduli space
of irreducible EH connections which is a smooth K\"{a}hler manifold.
The expectation value,
\eqn\apata{
<\exp(\tw + \e\Theta)> =
\sum_{r,s}\Fr{1}{\ell!s!}\e^s<\tw^\ell\,\Theta^s>,
}
can be interpreted as the intersection pairings on the moduli
space of $\o$-stable bundles for general $n$.
The right hand side of \apata\ is a non-vanishing invariant
if
\eqn\iii{
\hbox{dim}_\msbm{C}(\CM)\equiv d = \ell + 2s,
}
which is originated from the $U$  number anomaly due to
the  $\bar\p$ zero modes.
{}From \aac\ and \qqa, one can identify $\tw$ with the K\"{a}hler form
on $\CM$ after the localization of the path integral, because
$F^{1,1}_A\wedge\o =0$ for an EH connection $A$.
Thus, the expectation value $<\exp \tw>$
reduces to the volume of the moduli space
\eqn\babo{
\left<\exp\tw\right>
=\Fr{1}{\#Z}\int_\CM\Fr{\tw^d}{d!}
=\Fr{1}{\#Z}\hbox{vol}(\CM),
}
provided that the moduli space is compact.
This implies that there is a close relation between the HYM and the
$N=2$ TYM theories.

\subsec{Mapping to $N=2$ HYM theory}

Witten showed that there is a simple mapping to physical YM
theory in two dimensions from the $N=1$ TYM theory \WittenD.
Using this equivalence,
he was able to find general expressions for the intersection
parings on the moduli space of flat connections.
To begin with, we briefly recall the basic strategy of Witten.
Let $L = -i\dw W$ be the original action of the $N=1$ TYM theory
in two dimensions.
Replacing $W$ by $W+t W^\pr$, we have TYM theory with action
\eqn\aaa{
L(t) = -i\dw(W + tW^\pr).
}
The deformed theory is equivalent to the original one $L(t=0)$
if  (i) $W^\pr$ is such that $L(t)$ has nondegenerate kinetic
energy for all $t$; (ii) the deformed theory does not have any
new fixed points to flow in from infinity\foot{
By the standard manipulation of
cohomological theory, one can also formally show that deformed theory is
independent to $t$ as long as  $t$ is non-zero}.
Witten chose of $W^\pr$ which does not obey the condition (ii)
such that the new fixed point locus is precisely the space of YM
connections. Then he compared the expectation values of certain
observables computed in the original and the deformed theories
by taking large imaginary $t$ limit.

In this subsection,
we will show that an analogous mapping exists from the $N=2$ TYM theory
to the $N=2$ HYM theory on a compact K\"{a}hler manifold.
Our basic observation is that that the localization
of $N=2$ TYM theory can be realized by following two steps; i) localization
to $\CA^{1,1}$ (not to $\CA^{1,1}/\CG$) by integrating
$H^{2,0}$, $H^{0,2}$, $\c^{2,0}$, and $\bar\c^{0,2}$ out, provided that
there are
no $\c^{2,0}$ and $\bar\c^{0,2}$ zero-modes.
ii) localization of $\CA^{1,1}$ to $\CM$
by integrating out $H^0,\c^0, B^0$, provided that there are no reducible
EH connections. The arena of the HYM theory is
precisely the intermediate state, $F^{2,0}=F^{0,2}=\Dp\p =\Dpp\bar\p=0$,
after the first localization. We will deform the
$N=2$ TYM theory to obtain a new $N=2$ TYM theory whose fixed point of locus
is the space of HYM connections. Since both theories have the same
$N=2$ supersymmetry and topological observables, we can compare
expectation values of observables evaluated in the two theories.
We, then, interpret the expectation value of a combination of observables
evaluated in the deformed theory as the partition function of the $N=2$
HYM theory.

\lin{Deformation to a New $N=2$ TYM theory}

We consider  one parameter family of the $N=2$ TYM theories
with the action
\eqn\aaa{
S(t) = \Fr{\bs\bbs -\bbs\bs}{2} (V + tV^\pr).
}
A suitable choice of $V^\pr$ is
\eqn\aab{
V^\pr = -\Fr{n}{h^2}\int_M \Fr{\o^n}{n!}\, \tr B^{0}B^{0}.
}
Note that the global $U$-number symmetry of the original $(t=0)$
theory is no longer maintained in the deformed $(t\neq 0)$ theory,
since $(U,R)$ numbers of $V^\pr$ are $(-4,0)$
so those of $\bs\bbs V^\pr$ are $(-2,0)$. The upshot is that
we can integrate $B_0$ out from the action,
\eqn\aacc{
S(t) =
\Fr{\bs\bbs-\bbs\bs}{2}
\biggl(-\Fr{1}{h^2}\int_M \tr( B^{2,0}\wedge * F^{0,2}
+ B^{0,2}\wedge * F^{0,2}) -\Fr{n}{h^2}\int_M \Fr{\o^n}{n!}\tr(
B_0 f + t B_0 B_0)\biggr),
}
leaving a local action
\eqn\aad{
S^\pr(t) = -\Fr{\bs\bbs-\bbs\bs}{2}
\biggl(\Fr{1}{h^2}\int_M \tr( B^{2,0}\wedge * F^{0,2}
+ B^{0,2}\wedge * F^{0,2}\biggl)
+ \Fr{\bs\bbs -\bs\bbs}{2}
\biggl(\Fr{n}{4h^2 t}\int_M \Fr{\o^n}{n!}\,\tr f^2\biggr).
}
We can check this explicitly by integrating out $H^0, B^0, \c^0$ and
$\bar\c^0$ from the action
\eqn\aae{\eqalign{
S(t) =
&\Fr{1}{h^2}\!\int_M\!\!\! \tr\biggl[
  -iH^{2,0}\wedge*F^{0,2}
  -iH^{0,2}\wedge*F^{2,0}
  -\fr{i}{2}[\w,B^{2,0}]\wedge * F^{0,2}
  +\fr{i}{2}[\w,B^{0,2}]\wedge * F^{2,0}
  \phantom{\biggl]}
  \cr
& +i\c^{2,0}\wedge *\Dpp\bar\p
  +i\bar\c^{0,2}\wedge *\Dp\p
  -\biggl( n iH^{0}(f +2t B^0)
  -i\bar\c^{0}\L \Dpp\p
  -i\c^{0}\L \Dp\bar\p
  \phantom{\biggl]}
  \cr
& +2n t\c^0\bar\c^0
  +\ha B^{0}\L\left((i\Dp \Dpp -i\Dpp\Dp)\w -2[\p,\bar\p]\right)
  \biggr)\Fr{\o^n}{n!}\biggr]\;,
  \cr
}}
which leads to
\eqn\aae{\eqalign{
S^\pr(t) =
&\Fr{1}{h^2}\!\int_M\!\!\! \tr\biggl[
  -iH^{2,0}\!\wedge*F^{0,2}
  -iH^{0,2}\!\wedge*F^{2,0}
  -\fr{i}{2}[\w,B^{2,0}]\wedge * F^{0,2}
  +\fr{i}{2}[\w,B^{0,2}]\wedge * F^{2,0}
  \phantom{\biggl]}
  \cr
& +i\c^{2,0}\!\wedge *\Dpp\bar\p
  +i\bar\c^{0,2}\!\wedge *\Dp\p
  +\Fr{1}{4t}\biggl(f\L ((i\Dp \Dpp -i\Dpp\Dp)\w -2[\p,\bar\p])
  \biggr)\Fr{\o^n}{n!}
  \cr
& +\Fr{1}{2nt}(\L\Dpp \p)(\L\Dp \bar\p)\Fr{\o^n}{n!}
  \biggr].
  \cr
}}
This is identical to \aad.

Now we examine what kind of localization governs the deformed
$N=2$ TYM theory. The $t$ independent part of $S^\pr(t)$
is a cohomological
theory which localize the theory to $T\CA^{1,1}$.
Integrations over $H^{2,0}$ and $H^{0,2}$ gives
delta function support to $F^{2,0} = F^{0,2} = 0$
and
the $\c^{2,0}$
and $\bar\c^{0,2}$ integrals show that
$\p$ and $\bar\p$ are tangent to $\CA^{1,1}$,
\eqn\lofes{
\Dp\p = 0, \qquad \Dpp\bar\p =0.
}
The $\w$ equation of motion in the $t$-dependent part
gives
\eqn\lanl{
(\Dpp ^*\Dpp + \Dp^*\Dp) f= \Da^*\Da f =0,
}
where we have used the K\"{a}hler identities.
Thus, we have
\eqn\afag{
0=\int_M \tr f*\Da^*\Da  f = \int \tr \Da f\wedge *\Da f
\longrightarrow \Da f = 0.
}
This is the classical equation of motion for
the HYM theory. Note that the $\p$ and $\bar\p$ equations of motion
are
\eqn\aaf{
\fr{1}{n}\Dp\L(\Dpp \p) + [\p,f] = 0 , \qquad
\fr{1}{n}\Dpp\L(\Dp \bar\p) + [\bar\p,f] = 0.
}

Thus, (the bosonic parts of) the locus $\CN \subset \CA^{1,1}/\CG$
of the fixed points of the deformed
theory is the disjoint unions of
the moduli space of EH connections ($f=0$) and the moduli space of the
higher critical points of HYM theory (reducible connections with $f\neq
0$). Note that the moduli space of EH connections is identical to
the symplectic quotient $\eufm{m}^{-1}(0)/\CG$.
On the other hand, the moduli space $U_\b$
of the higher critical points
with a constant value of $f_\b$  is not isomorphic to
the symplectic quotient $\eufm{m}^{-1}(f_\b \o^n/4\pi^2(n-1)!)/\CG = \CM_\b$.
The space $U_\b$ is at most a set of finite points, since
open dense subset of $\CM_\b$ consists of irreducible connections.
Put it differently, the fixed points locus $\CN$ consists of
the moduli space $\CM^*$ of irreducible EH connections and the gauge
equivalence classes of reducible connections in $\CA^{1,1}$.

\lin{Mapping to $N=2$ HYM theory}

Now we have a new $N=2$ TYM theory with the enlarged fixed point locus
$\CN$. We can study the expectation values of the topological observables
evaluated in this new theory and compare them with those of
the original one. With a suitable choice of a set of observables, we may
be able to cleverly extract the contribution of $\CM$ from the path
integral performed over $\CN$. The observables $\Theta,\tw$ and
$\widetilde{c}_2$ are very special among the set of all observables.
They are nontrivial and non-degenerate because
$\o$ is nowhere vanishing $(1,1)$-form in the non-trivial second cohomology.
We consider a expectation value
$<\exp(\tilde\o +\e\Theta - \widetilde{c}_2/\e)>^\pr$ of the deformed theory,
\eqn\aaj{\eqalign{
<\exp&(\tilde\o +\e\Theta - \widetilde{c}_2/\e)>^\pr\cr
 = &\Fr{1}{\hbox{vol}(\CG)} \int_{\CA}
 \CD A^\pr\,\CD A^\ppr\,
 \CD\p\,\CD\bar\p\,\CD\w\,
 \CD  B^{2,0}\,\CD B^{0,2}\,
 \CD  \c^{2,0}\,\CD \bar\c^{0,2}\,
 \CD  H^{2,0}\,\CD H^{0,2}
 \cr
 &\times \exp\biggl[
 (\bs\bbs-\bbs\bs)
     \left(-\Fr{n}{8h^2 t}\int_M\!\Fr{\o^n}{n!}\tr f^2
 +\Fr{1}{2h^2}\!\int_M\! \tr\left(B^{2,0}\wedge * F^{0,2} +
 B^{0,2}\wedge * F^{2,0}\right)\right)
 \cr
 &+\Fr{1}{4\pi^2}\int_M \tr\bigl(i\w F^{1,1}
 +\p\wedge\bar\p \bigr)\wedge\Fr{\o^{n-1}}{(n-1)!}
 +\Fr{\e}{8\pi^2}\int_M \Fr{\o^n}{n!}\,\tr\w^2
 \cr
 &-\Fr{1}{4\pi^2\e}\int_M \tr(F^{2,0}\wedge F^{0,2}
 +\ha F^{1,1}\wedge F^{1,1})\wedge \Fr{\o^{n-2}}{(n-2)!}
\biggr],
\;\cr
}}
The above correlation function is formally independent of $t$ for
$t\neq 0$.
It is really independent of $t$ even at $t=\infty$ as long as,
in varying $t$,
the Lagrangian remains non-degenerate and with a good behavior at infinity
in field space \WittenD. The above choice ensures that these conditions are
obeyed as in the two-dimensional ($n=1$) case.
Thus we can simply discard the $u$-dependent part in \aaj\
by setting $t=\infty$. Then, the expectation value
$<\exp(\tilde\o +\e\Theta - \widetilde{c}_2/\e)>^\pr$ can be interpreted as
the partition function $Z(\e,\widetilde{c}_2)$ of a
$N=2$ supersymmetric theory
defined by the action functional
\eqn\aak{\eqalign{
I
=&\Fr{1}{h^2}\!\int_M\!\!\! \tr\biggl[
  -iH^{2,0}\wedge*F^{0,2}
  -iH^{0,2}\wedge*F^{2,0}
  -\fr{i}{2}[\w,B^{2,0}]\wedge * F^{0,2}
  +\fr{i}{2}[\w,B^{0,2}]\wedge * F^{2,0}
  \phantom{\biggl]}
  \cr
 &+i\c^{2,0}\wedge *\Dpp\bar\p
  +i\bar\c^{0,2}\wedge *\Dp\p
  \biggr]
  -\Fr{1}{4\pi^2}\int_M \tr\bigl(i\w F^{1,1}
    +\p\wedge\bar\p \bigr)\wedge\Fr{\o^{n-1}}{(n-1)!}
  \cr
 &-\Fr{\e}{8\pi^2}\int_M \Fr{\o^n}{n!}\,\tr\w^2
  +\Fr{1}{4\pi^2\e}\int_M \tr(F^{2,0}\wedge F^{0,2}
    +\ha F^{1,1}\wedge F^{1,1})\wedge \Fr{\o^{n-2}}{(n-2)!}
  \biggr].
  \cr
}}
It will be useful to show the equivalence of $Z(\e,\widetilde{c}_2)$
and the expectation value \aaj\ explicitly.
Since  both theories have the same $N=2$ supersymmetry,
it is sufficient to check that they have the same fixed points.
Provided that the localization of the both theory to
$T\CA^{1,1}$ is understood, we find the fixed point locus
of the latter theory is
$$
\left\{
\eqalign{
&\bs\p = \Dp \w,\cr
&\bbs\bar\p = \Dpp \w,\cr
&in f +\e\w = 0,\cr
}
\right.
\Longrightarrow
\left\{
\eqalign{
&\Dp f = 0,\cr
&\Dpp f = 0,\cr
}\right.
\Longrightarrow
\left\{
\eqalign{
&\bs \Dp f = -\fr{1}{n}\Dp\L(\Dpp\p) -[\p,f] =0,\cr
&\bbs \Dpp f = -\fr{1}{n}\Dpp\L(\Dp\bar\p) -[\bar\p,f] =0,\cr
}
\right.
$$
which are identical to \afag\ and \aaf.

The $N=2$ supersymmetric theory
defined by the action functional \aak\ (the $N=2$ HYM theory)
is the desired quantum field theoretical setting of the HYM
theory discussed in the previous section.  Before showing this,
we should comment that the $N=2$ HYM theory is well defined
as a quantum field theory. Obviously, It is a  finite
theory in the standard criteria, since its partition function
$Z(\e,\tilde{c}_2)$ is identical to the expectation value
$<\exp(\tilde\o +\e\Theta - \widetilde{c}_2/\e)>^\pr$ of
the topological observables evaluated in a cohomological theory.
The subtle point is that even the cohomological field
theory has difficulties in dealing with the non-compactness
and the singularities in the moduli space.
The path
integral of the $N=2$ HYM theory has two kinds of contributions,
from the moduli space $\CM$ of EH connections and from the higher
critical points $U_\b$. Even though the path integral contributed from
$\CM$ is identical to the path integral of the original
$(t =0)$ $N=2$ TYM theory, a new problem may appear due to the
contributions of higher critical points. However, the contributions
of $U_\a$ can be precisely determined, since $U_\a$ consists of,
at most, finite collection of points.

We also have  a partial resolution
of the singularities in the moduli space.  Since we have already
integrated out $\c^0$ and $\bar\c^0$, there are no fermionic zero-modes
due to reducible EH connections. The (bosonic)
$\w$-zero-modes due to reducible EH connections is not so harmful
because of the $\tr\w^2$ term in the action.
\lin{Final Reduction}

We assume that the cohomology  $H^2(\hbox{End}_0(\CE_A))$
is trivial everywhere.
Then, we can integrate
$H^{2,0}$, $H^{0,2}, B^{2,0}, B^{0,2}, \c^{2,0}$
and $\bar\c^{0,2}$ out and the partition function $Z(\e, \tilde{c}_2)$
of the $N=2$ HYM theory becomes
\eqn\aal{\eqalign{
\exp(-\Fr{c_2}{\e})\times
\Fr{1}{\hbox{vol}(\CG)}\int_{\CA}
\!\!\CD \! A^\pr\,\CD \! A^\ppr\,
\CD \p\,\CD \bar\p\,\CD \w\cdot\!
\prod_{x\in M}
\d(F^{2,0}(x))
\d(F^{0,2}(x))
\d(\Dp\p(x))
\d(\Dpp\bar\p(x))
\cr
\times\exp\biggl(
\Fr{1}{4\pi^2}\int_M \tr\bigl(i\w F^{1,1} +
\p\wedge\bar\p \bigr)\wedge\Fr{\o^{n-1}}{(n-1)!}
+\Fr{\e}{8\pi^2}\int_M \Fr{\o^n}{n!}\tr\w^2\biggr).\cr
}}
The measure of the above path integral is
equivalent to
\eqn\mmema{
\int_{\CA}
\!\!\CD \! A^\pr\,\CD \! A^\ppr\, \CD \p\,\CD \bar\p\,
\CD \w\cdot\! \prod_x
\d(F^{2,0})
\d(F^{0,2})
\d(\Dp\p)
\d(\Dpp\bar\p) \cdots\equiv
\int_{T\CA^{1,1}}
\!\!\!\!\CD \! A\,\CD \! A^\ppr\, \CD \p\,\CD \bar\p\,
\CD \w\cdots ,
}
since there are no loop corrections of the delta function
constraints,
$\prod\d(F^{2,0}(x))\d(F^{0,2}(x))$ and
$\prod\d(\Dp\p(x))\d(\Dpp\bar\p(x))$, due to the $N=2$ fermionic
symmetry,
\eqn\illust{
\bs F^{2,0} = \Dp\p, \qquad
\bbs F^{2,0} =0, \qquad
\bbs F^{0,2} = \Dpp\bar\p, \qquad
\bs  F^{0,2} =0.
}
Thus, the partition function \aal\ can be written as
\eqn\aam{\eqalign{
\exp(-\Fr{c_2}{\e})\times
\Fr{1}{\hbox{vol}(\CG)}&\int_{T\CA^{1,1}}
\!\!\CD \! A^\pr\,\CD \! A^\ppr\,
\CD \p\,\CD \bar\p\,\CD \w\cr
&\exp\biggl(
\Fr{1}{4\pi^2}\int_M \tr\bigl(i\w F^{1,1} +
\p\wedge\bar\p \bigr)\wedge\Fr{\o^{n-1}}{(n-1)!}
+\Fr{\e}{8\pi^2}\int_M \Fr{\o^n}{n!}\tr\w^2\biggr).\cr
}
}
This can be viewed as the partition function of the HYM theory
after integrating $\w$ out. The role of the decoupled fields
$\p$ and $\bar\p$ is to give a symplectic (K\"{a}hler) measure
on $\CA^{1,1}$, which will be denoted as $\CD A$,
and to ensure the $N=2$ supersymmetry.
Now we can extract one important conclusion that
the partition function of the HYM theory can be expressed as a
sum of contribution of critical points $\CN$
\eqn\huj{
Z(\e,c_2) = \exp(-c_2/\e)\times\sum_{\a\in \CN}Z_\a (\e).
}
This coincides to the prediction of the Witten's non-Abelian
localization theorem.
In sect.~3.1, we showed that $<\exp\tw>$ can be identified with
the volume of $\CM$ under suitable topological conditions.
{}From \aaj\ and \aam,  one can write $Z(0) = <\exp\tw>^\prime$. The
$\w$ integral gives delta function support on EH connections and
there are no additional critical points contributing to the path integral.
This leads that the deformed theory is equivalent to the original
theory in this particular case.
Thus, we can recover \tyh\ from \babo,
\eqn\aal{
<\exp \tilde\o>^\pr = <\exp \tilde\o>
= \Fr{1}{\# Z}\hbox{vol}(\CM).
}

Now the key step is to find the relations between the original $N=2$ TYM and
the $N=2$ HYM
theories in general situation, $\e\neq 0$.
By integrating $\w$ and $\p,\bar\p$ out in \aam, we have
\eqn\aan{
\left<\exp(\tilde\o +\e\Theta -\tilde c_2/\e)\right>^\pr =
\exp(-\Fr{c_2}{\e})\times
\Fr{1}{\hbox{vol}(\CG)}\int_{\CA^{1,1}}
\!\!\CD \! A\,\exp\biggl(
\Fr{n}{8\pi^2\e}\int_M \Fr{\o^n}{n!}\, \tr f^2 \biggr).
}
Now,
the two expectation values
$<\exp(\tilde\o +\e\Theta -\tilde c_2/\e)>$ and
$<\exp(\tilde\o +\e\Theta -\tilde c_2/\e)>^\pr$ are no longer
identical. The later has contributions from the higher critical
points $\left\{U_\b\right\}$. Since a $U_\b$ consists of
finite set of points, the relation \aan\ clearly shows that their
contributions are exponentially small, involving the relevant values of
the action of the HYM. Thus
\eqn\aao{\eqalign{
\exp(-\Fr{c_2}{\e})\times
\Fr{1}{\hbox{vol}(\CG)}&\int_{\CA^{1,1}}\!\!\CD \! A
\exp\left(\Fr{n}{8\pi^2\e}\int_M\Fr{\o^n}{n!}\, \tr f^2 \right)\cr
&=\left<\exp(\tilde\o +\e\Theta -\tilde c_2/\e)\right>
+ O(\exp(-c_2/\e - c/\e)),\cr
}
}
where $c$ is the smallest value of
$\e I(A) = -\fr{n}{4\pi^2}\int_M \Fr{\o^n}{n!}\, \tr f^2$ at the higher
critical points.

\lin{Higher Critical Points}

The critical points of the HYM theory are the solutions of the classical
equations of motion $\Da f =0$.
One obvious solution is $f = 0$ which corresponds to
the EH connections. If there is a higher critical point $f\neq 0$,
it means that the connection is reducible.
For simplicity, we consider the $SU(2)$ case only.
We further assume that $M$ is a projective algebraic
manifold\foot{Note that every compact Riemann surface is
projective algebraic.}
such that we can pick a Hodge metric and associated
K\"{a}hler form $\o \in H^{1,1}(M,\BZ)$.

If there is a reducible $SU(2)$ connection
$A\in \CA^{1,1}$,
the associated holomorphic vector
bundle $\CE_A$ is decomposed into the direct sum of
$U(1)$ holomorphic line bundles,
\eqn\kkn{
\CE_A = \CU\oplus \CU^{-1},
}
whose curvature $F_A \in \O^{1,1}(\hbox{End}_0(\CE_A)$ is constant
in each higher critical point,
\eqn\kko{
F_A =
\left(\matrix{&F_c&0\cr
&0&-F_c\cr}\right) \in \eufm{su(2)}.
}
A holomorphic line bundle $\CU$ is classified by its first Chern class
$c_1(\CU)$
\eqn\kks{
c_1(\CU) =\Fr{i}{2\pi} F_c\;\;\;\; \in H^{1,1}(M,\BZ).
}
We can decompose $c_1(\CU)$ as
\eqn\kkv{
\Fr{i}{2\pi}F_c = \Fr{i}{2\pi}f_c\o + \Fr{i}{2\pi}S,\qquad S\wedge
\o^{n-1} =0,
}
where the part orthogonal to $\o$ is an element of the primitive
cohomology class $P^{1,1}(M,\BZ)$.
The degree of $\CU$,  deg($\CU$), is defined by
\eqn\kku{
\hbox{ deg}(\CU) =\int_M c_1(\CU)\wedge \o^{n-1}
= \Fr{i}{2\pi}\int_M F_c\wedge \o^{n-1}
= \Fr{if_c}{2\pi}\int_M \o^n \in \BZ,
}
The {deg}($\CU$) is a (real) topological invariant
depending only on $c_1(\CU)$ and the de Rahm cohomology class of $\o$.
The constant value $f_c$ is
determined by
\eqn\kkw{
\Fr{if_c}{2\pi} = \Fr{\hbox{deg}(\CU)}{\int_M \o^n}.
}
If we consider the $n=1$ case, the first Chern number $<c_1(\CU), M>$
is identical to {deg}($\CU$) which can be an arbitrary integer.
This is sufficient to
determine all of the allowed values of $f_c$.

For $n > 1$, we have another important condition
\eqn\kkp{
<c_2(\CE_A)\smile \o^{n-2},[M]> = - <c_1(\CU)\smile c_1(\CU)\smile
\o^{n-2},[M]>,
}
that is
\eqn\kkr{\eqalign{
(n-2)!c_2&= \Fr{1}{8\pi^2}\int_M \tr (F_A\wedge F_A)\wedge \o^{n-2}\cr
& = -\int_M
c_1(\CU)\wedge c_1(\CU)\wedge \o^{n-2}\cr
& = -\int_M (\Fr{i}{2\pi}f_c)(\Fr{i}{2\pi}f_c)\o^n
 - \int_M \Fr{i}{2\pi}S\wedge \Fr{i}{2\pi} S \wedge \o^{n-2}\in \BZ^+. \cr
}
}
This is precisely the Hodge-Riemann bilinear form $Q$
among $H^{1,1}(M,\BZ)$ \GH,
\eqn\bilinear{
Q(\eta,\eta) = \int_M \eta\wedge \eta \wedge \o^{n-2} = -(n-2)!c_2.
}
Thus, we should find the every solution
$\pm \eta \in H^{1,1}(M,\BZ)$ of the bilinear form to determine
the values $f_c$ at the critical points. If $c_1(\CU)$ is a solution,
the value of $I(A)$ is given by
\eqn\accct{
I(A) = -\Fr{n^2}{8\pi^2\e}\int_M \Fr{\o^n}{n!}\tr f^2
     = \Fr{n}{\e (n-1)!}\Fr{(\hbox{deg}(\CU))^2}{\int \o^n}
     = \Fr{n}{\e (n-1)!} m^2
}
where $m$ is a certain integer which can be determined if we can
solve \bilinear. Eq.\kkr\ shows that a reducible EH connection is a
solution of
\eqn\krk{
Q(\eta_\perp,\eta_\perp) =  -(n-2)!c_2,
}
where $\eta_\perp \in P^{1,1}(M,\BZ)$. From the Hodge-Riemann
bilinear relations, we have
\eqn\hodger{
(-1)^{(2n-2)(2n-3)/2}Q(\eta_\perp,\eta_\perp) > 0 .
}
Since $c_2 > 0$ to admit the EH connections, we can conclude
that there are no reducible EH connections if $(2n-2)(2n-3)/2$
is an non-zero even integer.

\lin{Application to the Donaldson Invariants}

The most interesting mathematical application of the $N=2$ HYM theory
is to the Donaldson polynomial invariants.
Let $S$ be a simply connected algebraic surface with
a Hodge metric and associated K\"{a}hler form $\o \in H^{1,1}(S,\BZ)$.
We further assume that the geometric genus $p_g(S) > 0$ is strictly
positive. Let $H$ be an algebraic cycle
Poincar\'{e} dual to $\o$. Then $H$ is
an ample divisor that some integer multiple of $H$ is the hyperplane
section of an suitable embbeding $S \subset \CP^m$.
Let $E$ be a (complex) vector
bundle over $S$ with reduction of structure group to $SU(2)$.
The bundle $E$ is classified by the instanton number $k=<c_2(E),S>$
which will be always assumed to be  strictly positive.
The topological observables $\tw$ and $\Theta$
define the Donaldson $\m$-maps
\eqn\cca{\eqalign{
\m : H_2(S,\BZ) \rightarrow H^{1,1}(\CA^*_k/\CG,\BZ),\cr
\m : H_0(S,\BZ) \rightarrow H^{2,2}(\CA^*_k/\CG,\BZ).\cr
}
}
The expectation value $<\exp (\tw + \e\Theta)>$ evaluated in the original
$N=2$ TYM theory corresponds to the Donaldson invariants
\eqn\ccb{
\eqalign{
<\exp (\tw + \e\Theta)> &=
\Fr{1}{\# Z}\sum_{r,s}^{r + 2s = d}\!\!\Fr{\e^s}{r! s!}
q_{k,S}(\overbrace{H,..,H}^r,\overbrace{pt,..,pt}^s)\cr
&=\Fr{1}{\# Z}\sum_{r,s}^{r + 2s = d}\!\!\Fr{\e^s}{r! s!}
<\m(H)\smile\cdots\smile\m(H)\smile\m(pt)
\cdots\smile\m(pt),[\CM_k]>.\cr
}}
where $d = 4k -3(1 + p_g(S))$ denotes the complex dimension
of the moduli space $\CM$ of anti-self-dual connections.
Mathematically, the Donaldson invariants $q_{k,S}(H,..H,pt,..pt)$
is well defined for $p_g(S) > 0$ and for large enough $k$.
The former condition is that there are no reducible instantons
and  the latter one ensures that the Hodge metric behaves as a generic
metric so the $H^2(\hbox{End}_0(\CE_A))$ cohomology is trivial.

The expectation value $<\exp (\tw + \e\Theta)>^\pr$
evaluated by the deformed $N=2$ TYM theory, which is identical to
\ccb\ up to exponentially small term, can be represented by
the partition function $Z(\e)_k$ of the $N=2$ HYM theory without
the topological term $k/\e$ in the action.
With the conditions stated above, we have
a reduction similar to \aao,
\eqn\twed{
Z(\e)_k =\Fr{1}{\hbox{vol}(\CG)}\int_{\CA^{1,1}_k}\!\!\CD \! A
\exp\left(\Fr{n}{8\pi^2\e}\int_M\Fr{\o^n}{n!}\, \tr f^2 \right)
=\left<\exp(\tilde\o +\e\Theta)\right> + O(\exp(- c/\e)).
}
Thus, the partition function $Z(\e)_k$ in a small $\e$ limit
becomes
\eqn\twwd{
Z(\e)_k  = \Fr{1}{\# Z}\sum_{r,s}^{r + 2s = d}\Fr{\e^s}{r! s!}
q_{k,S}(\overbrace{H,..,H}^r,\overbrace{pt,..,pt}^s) + \hbox{
exponentially small terms}.
}

\newsec{Further Studies}

We have seen that the $N=2$ TYM theory is equivalent to the $N=2$
HYM theory. Using this equivalence, we can
obtain general expressions of certain intersection parings
on the moduli space of EH connections on a compact
K\"{a}hler manifold by solving the
$N=2$ HYM theory. The question is whether we can evaluate the
partition function exactly even for $n > 1$ cases.
We may use the fixed point theorem of Witten \WittenE\ in evaluating
the partition function, since the $N=2$ HYM theory has global fermionic
symmetry. To do actual calculation, we should be able to find
every HYM connections. However, this will be difficult in general
for the $n >2$ cases. Furthermore, we do not understand many properties
of the moduli space of EH connections in those cases.
The most tractable case is the $N=2$ HYM theory on
algebraic surfaces with the structure group $SU(2)$ \HP.

We have also seen that the $N=2$ HYM theory is a natural higher dimensional
analogue of
physical Yang-Mills theory in two dimensions.
It will be interesting
to examine whether other noble properties of the two dimensional
Yang-Mills theory not studied in this paper remain valid in the $N=2$
HYM theory;
i) It is known that the action functional of
Yang-Mills theory in two dimensions is an equivariantly perfect Morse
functional \AB. This is almost followed from the fact
that the action functional is the norm squared
of the moment map \Kirwan. Thus, we can expect that the action functional
of the HYM theory is a perfect Morse functional even for the higher
dimensional cases.
Then, it will be possible to obtain the Poincar\'{e} polynomials
of the moduli space of EH connections.
ii) Recently, Gross et.~al.~showed that two dimensional
Yang-Mills theory has a simple interpretation as a string theory\Gross.
It will be interesting to explore the possible stringy behavior
of the $N=2$ HYM theory, which will be much easier than the case of physical
Yang-Mills theory.

\ack{I am grateful to H.J.~Kim and Q.-H.~Park for
useful discussions.
I would like to thank G.~Thompson for his suggestions
and criticisms after reading the first version of this work.
I would also like to thank the referee of this paper for valuable comments.
It is my pleasure to express my gratitude to S.~Hyun for many exiting
discussions and reading on this paper.}

\listrefs
\end